\documentclass[reprint,superscriptaddress,amsmath,amssymb,
aps,prl,10pt, twocolumn,floatfix]{revtex4-2}

\usepackage{graphicx}
\usepackage{dcolumn}
\usepackage{bm}
\usepackage{placeins}
\usepackage[caption=false]{subfig}
\usepackage{multirow}

\usepackage[colorlinks=true,breaklinks=true]{hyperref}
\hypersetup{allcolors=[rgb]{0.0 0.0 1.0},linkcolor=[rgb]{0.75 0.05 0.05}}
\usepackage[dvipsnames]{xcolor}

\usepackage{booktabs} 
\AtBeginDocument{
\heavyrulewidth=.08em
\lightrulewidth=.05em
\cmidrulewidth=.03em
\belowrulesep=.65ex
\belowbottomsep=0pt
\aboverulesep=.4ex
\abovetopsep=0pt
\cmidrulesep=\doublerulesep
\cmidrulekern=.5em
\defaultaddspace=.5em
}

\usepackage{siunitx}
\DeclareSIUnit\eVperc{\eV\per\clight}
\DeclareSIUnit\clight{\text{\ensuremath{c}}}
\usepackage{physics}

\def\MADMAX       {\mbox{{\sc Madmax}}}

\begin{document}


\title{First search for dark photon dark matter with a \MADMAX{} prototype}

\author{J.~Egge}
\email[Correspondence to: ]{jacob.egge@uni-hamburg.de}
\affiliation{Universität Hamburg, Hamburg, Germany}

\author{D.~Leppla-Weber}
\affiliation{Deutsches Elektronen-Synchrotron DESY, Germany}

\author{S.~Knirck}
\affiliation{Fermi National Accelerator Laboratory, Batavia, USA}

\author{B.~Ary dos Santos Garcia}
\affiliation{III. Physikalisches Institut A,  RWTH Aachen University, Aachen, Germany}

\author{D.~Bergermann}
\affiliation{III. Physikalisches Institut A,  RWTH Aachen University, Aachen, Germany}

\author{A.~Caldwell}
\affiliation{Max-Planck-Institut f{\"ur} Physik, Garching, Germany}

\author{V.~Dabhi}
\affiliation{Centre de Physique des Particules de Marseille, Aix Marseille Univ, CNRS/IN2P3, CPPM, Marseille, France}

\author{C.~Diaconu}
\affiliation{Centre de Physique des Particules de Marseille, Aix Marseille Univ, CNRS/IN2P3, CPPM, Marseille, France}

\author{J.~Diehl}
\affiliation{Max-Planck-Institut f{\"ur} Physik, Garching, Germany}

\author{G.~Dvali}
\affiliation{Max-Planck-Institut f{\"ur} Physik, Garching, Germany}

\author{M.~Ekmedžić}
\affiliation{Universität Hamburg, Hamburg, Germany}

\author{F.~Gallo}
\affiliation{Centre de Physique des Particules de Marseille, Aix Marseille Univ, CNRS/IN2P3, CPPM, Marseille, France}

\author{E.~Garutti}
\affiliation{Universität Hamburg, Hamburg, Germany}

\author{S.~Heyminck}
\affiliation{Max-Planck-Institut f{\"ur} Radioastronomie, Bonn, Germany}

\author{F.~Hubaut}
\affiliation{Centre de Physique des Particules de Marseille, Aix Marseille Univ, CNRS/IN2P3, CPPM, Marseille, France}

\author{A.~Ivanov}
\affiliation{Max-Planck-Institut f{\"ur} Physik, Garching, Germany}

\author{J.~Jochum}
\affiliation{Physikalisches Institut, Eberhard Karls Universit{\"a}t T{\"u}bingen, T{\"u}bingen, Germany}

\author{P.~Karst}
\affiliation{Centre de Physique des Particules de Marseille, Aix Marseille Univ, CNRS/IN2P3, CPPM, Marseille, France}

\author{M.~Kramer}
\affiliation{Max-Planck-Institut f{\"ur} Radioastronomie, Bonn, Germany}

\author{D.~Kreikemeyer-Lorenzo}
\affiliation{Max-Planck-Institut f{\"ur} Physik, Garching, Germany}

\author{C.~Krieger}
\affiliation{Universität Hamburg, Hamburg, Germany}

\author{C.~Lee}
\affiliation{Max-Planck-Institut f{\"ur} Physik, Garching, Germany}


\author{A.~Lindner}
\affiliation{Deutsches Elektronen-Synchrotron DESY, Germany}

\author{J.~P.~A.~Maldonado}
\affiliation{Max-Planck-Institut f{\"ur} Physik, Garching, Germany}

\author{B.~Majorovits}
\affiliation{Max-Planck-Institut f{\"ur} Physik, Garching, Germany}

\author{S.~Martens}
\affiliation{Universität Hamburg, Hamburg, Germany}

\author{A.~Martini}
\affiliation{Deutsches Elektronen-Synchrotron DESY, Germany}

\author{A.~Miyazaki}
\affiliation{Université Paris-Saclay, CNRS/IN2P3, IJCLab, Orsay, France}

\author{E.~Öz}
\affiliation{III. Physikalisches Institut A,  RWTH Aachen University, Aachen, Germany}

\author{P.~Pralavorio}
\affiliation{Centre de Physique des Particules de Marseille, Aix Marseille Univ, CNRS/IN2P3, CPPM, Marseille, France}

\author{G.~Raffelt}
\affiliation{Max-Planck-Institut f{\"ur} Physik, Garching, Germany}

\author{A.~Ringwald}
\affiliation{Deutsches Elektronen-Synchrotron DESY, Germany}

\author{J.~Redondo}
\affiliation{Universidad Zaragoza, Zaragoza, Spain}

\author{S.~Roset}
\affiliation{Centre de Physique des Particules de Marseille, Aix Marseille Univ, CNRS/IN2P3, CPPM, Marseille, France}

\author{N.~Salama}
\affiliation{Universität Hamburg, Hamburg, Germany}

\author{J.~Schaffran}
\affiliation{Deutsches Elektronen-Synchrotron DESY, Germany}

\author{A.~Schmidt}
\affiliation{III. Physikalisches Institut A,  RWTH Aachen University, Aachen, Germany}

\author{F.~Steffen}
\affiliation{Max-Planck-Institut f{\"ur} Physik, Garching, Germany}

\author{C.~Strandhagen}
\affiliation{Physikalisches Institut, Eberhard Karls Universit{\"a}t T{\"u}bingen, T{\"u}bingen, Germany}

\author{I.~Usherov}
\affiliation{Physikalisches Institut, Eberhard Karls Universit{\"a}t T{\"u}bingen, T{\"u}bingen, Germany}

\author{H.~Wang}
\affiliation{III. Physikalisches Institut A,  RWTH Aachen University, Aachen, Germany}

\author{G.~Wieching}
\affiliation{Max-Planck-Institut f{\"ur} Radioastronomie, Bonn, Germany}

\collaboration{\MADMAX{} Collaboration}
\noaffiliation

\author{G.~Cancelo}
\affiliation{Fermi National Accelerator Laboratory, Batavia, USA}

\author{M.~Di Federico}
\altaffiliation[Also at: ]{Universidad Nacional del Sur, IIIE-CONICET, Argentina}
\affiliation{Fermi National Accelerator Laboratory, Batavia, USA}

\author{G.~Hoshino}
\affiliation{Department of Physics, University of Chicago, Chicago, USA}

\author{L.~Stefanazzi}
\affiliation{Fermi National Accelerator Laboratory, Batavia, USA}

\date{August 5, 2024}

\begin{abstract}
We report the first result from a dark photon dark matter search in the mass range from $\SIrange{78.62}{83.95}{\micro\eVperc^2}$ with a dielectric haloscope prototype for \MADMAX{} (Magnetized Disc and Mirror Axion eXperiment). Putative dark photons would convert to detectable photons within a stack consisting of three sapphire disks and a mirror. The emitted power of this system is received by an antenna and successively digitized using a low-noise receiver. No significant signal attributable to dark photons has been observed above the expected background. Assuming unpolarized dark photon dark matter with a local density of $\rho_{\chi}=\SI{0.3}{\giga\eV\per\cm^3}$ we exclude a dark photon to photon mixing parameter $\chi > 2.7 \times 10^{-12}$ over the full mass range and $\chi > 1.1 \times 10^{-13}$ at a mass of $\SI{80.57}{\micro\eVperc^2}$ with a 95\% confidence level. This is the first physics result from a \MADMAX{} prototype and exceeds previous constraints on $\chi$ in this mass range by up to almost three orders of magnitude.
\end{abstract}

\maketitle

{\bf\textit{Introduction.}}---The nature of dark matter (DM) might be the greatest unsolved mystery of particle physics and cosmology. Many experiments try to directly detect DM particles from the galactic halo in the laboratory. In recent times, very light-weight DM candidates with masses below $\sim\SI{1}{\milli\eVperc^2}$ have received increased attention~\cite{Billard:2021uyg}. One such wave-like DM candidate is the dark photon (DP), also known as hidden photon~\cite{darkphoton_handbook,dark_photon_primer}. DPs appear in extensions of the Standard Model (SM) that postulate an additional $U_{\chi}(1)$ gauge symmetry. If this symmetry is broken, the DP gains a mass $m_{\chi}$. Assuming that the SM fields remain uncharged under the new $U_{\chi}(1)$, DPs would predominantly interact with SM particles via kinetic mixing with the ordinary photon. Its mass and small interaction strength with SM fields makes the DP a suitable DM candidate. Dark photon dark matter (DPDM) production from inflationary perturbations with $m_{\chi}\sim\SI{100}{\micro\eVperc^2}$ would easily saturate the observations ~\cite{Graham_dp_infl,Nakai2020_dp_infl} and its discovery could pinpoint the scale of inflation. In the low energy limit, the interaction between DPs and photons can be described by additional source terms in the classical Maxwell's equations. In particular, Ampère's law is modified to~\cite{Huang:2018mkk,marocco2021dark}
\begin{equation}
    \curl{\vb*{H}} - \vb*{\dot{D}} = \chi \epsilon_0 \qty(\vb*{\dot{E}}_\chi - c^2 \curl{\vb*{B}_\chi}),
    \label{Eq:ampere_dark_photon}
\end{equation}
where $(\vb*{E}_{\chi}, \vb*{B}_{\chi})$ are the DP electric and magnetic fields, $\chi$ is the kinetic mixing angle, and $(\vb*{H},\vb*{D})$ the magnetic and displacement field of ordinary electrodynamics. For DPDM, the spatial derivative $\curl{\vb*{B}_\chi}$ can be neglected as the de Broglie wavelength is much larger than the setup~\cite{MADAMAX_vel_2017,MADMAX_directinal_2018} whereas the temporal derivative $\vb*{\dot{E}}_\chi$ acts as an effective current density that oscillates with frequency $\nu_{\chi}\approx {m_{\chi}c^2}/{h}$. The oscillating DP electric field drives the free charges inside a conductor which leads to emissions of ordinary photons perpendicular to its boundary. The power emitted of a metallic mirror is proportional to its area $A$ and the local DPDM density $\rho_{\chi}=\frac{\epsilon_0}{2} \abs{\vb*{E}_\chi}^2$~\cite{Horns:2012jf},
\begin{equation}
    P_0 = \chi^2  c \rho_{\chi}  A  \alpha_{\rm{pol}}^2.
    \label{Eq:power_dish}
\end{equation}
The factor $\alpha_{\rm{pol}}$ accounts for the DP polarization and is the average fraction of DPs that the experiment is sensitive to. We assume unpolarized DPDM, setting $\alpha_{\rm{pol}}^2=1/3$ as we are sensitive only to a single polarization~\cite{darkphoton_handbook}. 

In this letter, a search for DPDM using a \MADMAX{} (MAgnetized Disk and Mirror Axion eXperiment) prototype setup is presented. \MADMAX{} is designed to search for axions and axion-like particles in an external static B-field in the mass range of \SIrange{40}{400}{\micro\eVperc^2}~\cite{MADMAX2017,MADMAX:2019pub,madmax_theo_found}. With no B-field applied, the setup is still sensitive to DPs. The proposed mass range is difficult to access using traditional designs like cavity haloscopes because the effective volume where DPs (or equivalently axions) convert into photons naively scales as $\lambda^3 \propto 1/m_{\chi}^3$. This makes large, and therefore sensitive, conversion volumes hard to achieve. The dielectric haloscope concept of \MADMAX{} removes the dependency of the conversion volume from the wavelength by placing dielectric disks in front of a metallic mirror~\cite{Jaeckel:2013eha}. This concept has already been successfully employed in searches for DPDM at lower~\cite{Cervantes:2022epl} and higher~\cite{Chiles:2021gxk} DP masses. The axion or DP-induced photon emission of the disks and mirror, collectively called the booster, can constructively interfere and resonate between disks. The overall increase in expected DP signal power w.r.t.~a single mirror is described by the boost factor $\beta^2 = \frac{P_{\rm{sig}}}{P_0}$, where $P_{\rm{sig}}$ is the signal power received by the first preamplifier of the receiver system. By controlling the spacing between disks, the boost factor can be tuned in frequency~\cite{MADMAX_PoP_2020} allowing for both broadband and resonant searches. This enables scanning for a broad range of possible DP masses. In this work, a single fixed booster configuration at room temperature without scanning capability is used. The expected sensitivity to $\chi$ from a DP signal with bandwidth $\Delta\nu_{\chi}\approx 10^{-6} \nu_{\chi}$ can be expressed via Dicke's radiometer equation~\cite{dicke1946} as
\begin{equation}\label{Eq:chi_parameters}
\begin{split}
    \chi = 1.0 \times 10^{-13} \left(\frac{640}{\beta^2}\right)^{1/2} \left(\frac{\SI{707}{\centi\meter^2}}{A}\right)^{1/2} \\
    \times\left(\frac{T_{\rm{sys}}}{\SI{240}{\kelvin}}\right)^{1/2} 
    \left(\frac{\SI{11.7}{\day}}{\Delta t}\right)^{1/4} 
    \left(\frac{\rm{SNR}}{5}\right)^{1/2} \\
    \times\left(\frac{\SI{0.3}{\giga\eV\per\centi\meter^3}}{\rho_{\chi}}\right)^{1/2} 
    \left(\frac{\Delta\nu_{\chi}}{\SI{20}{\kilo\hertz}}\right)^{1/4},
\end{split}
\end{equation}
where $T_{\rm{sys}}$ is the system noise temperature, $\Delta t$ the effective data-taking time, and $\rm{SNR}$ is the Signal to Noise Ratio of a hypothetical DP signal. Here, DPs are considered to comprise all of the local dark matter density assuming $\rho_{\chi}=\SI{0.3}{\giga\eV\per\cm^3}$~\cite{localCDM}. Both $\beta^2$ and $T_{\rm{sys}}$ are functions of frequency that depend on the booster configuration.

\begin{figure}
    \centering
    \includegraphics[width=\linewidth]{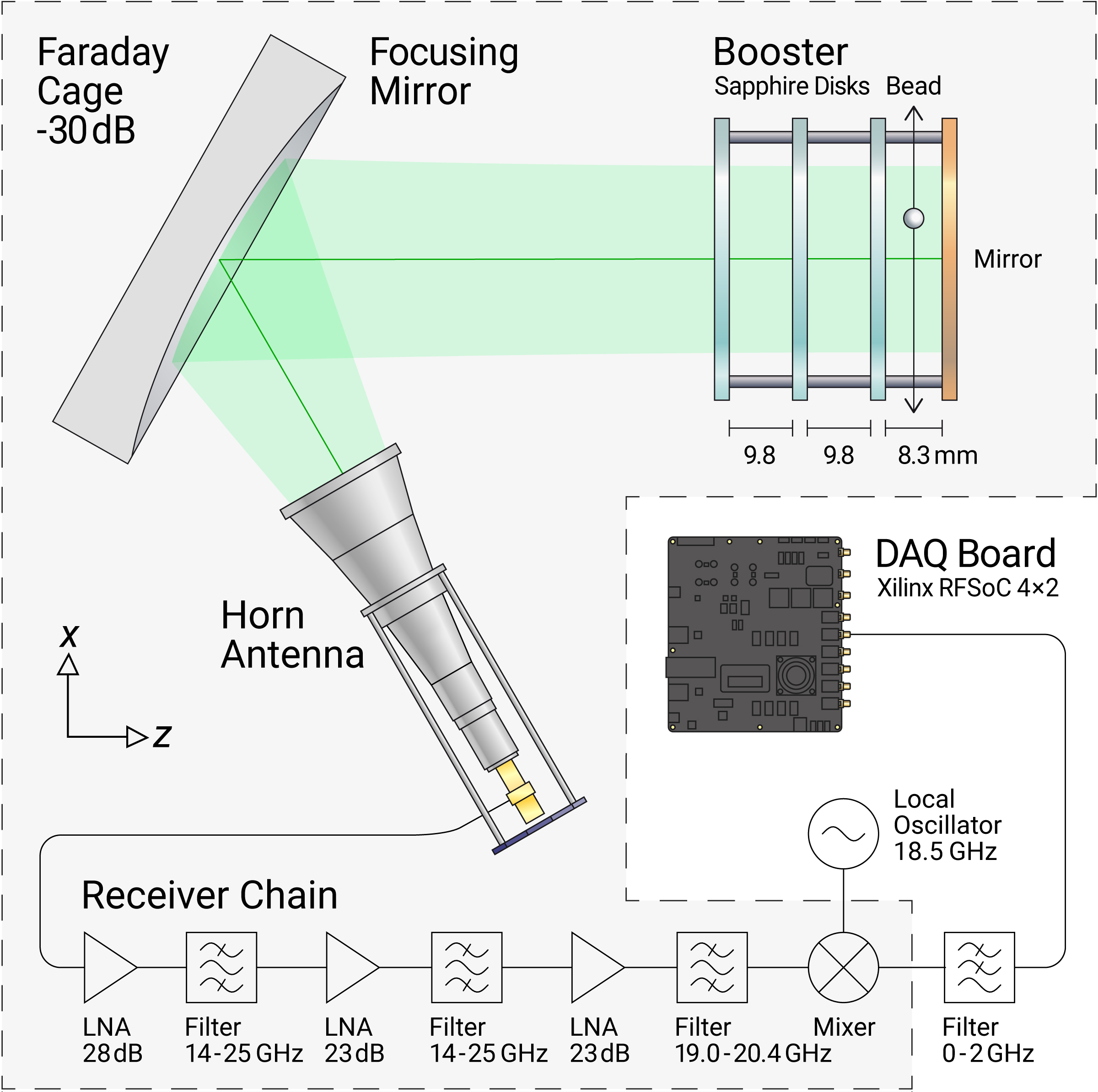}
    \caption{Experimental Setup. It is placed in an RF-isolated Faraday cage indicated by the dashed line. To determine the boost factor, a bead can be inserted into the booster to measure the electric field induced by a reflection measurement. Sketch not to scale.}
    \label{fig:setup}
\end{figure}

{\bf\textit{Experimental Setup.}}---A schematic of the setup is shown in Fig.~\ref{fig:setup}. Three sapphire disks with relative permittivity $\epsilon=9.3(1)$ and thickness $\SI{1.00(2)}{\milli\meter}$ as well as an aluminum mirror make up the booster of the dielectric haloscope. The disks and mirror have a diameter of $\SI{30}{\centi\metre}$ each and are held parallel to each other by mechanical spacers in a fixed configuration. The planarity of the three disks has been measured with the methodology presented in~\cite{madmax_p200} and gives an RMS of $\SI{50}{\micro\meter}$ ($\SI{200}{\micro\meter}$ min-max). The DP-induced emissions are coupled to a K-band Gaussian beam horn antenna via an off-axis ellipsoidal mirror of focal length $f=\SI{504}{\milli\metre}$. The horn antenna's position and orientation are precisely controlled using motorized stages. A low-noise amplifier (LNA) is connected directly to the horn antenna. Additional LNAs and bandpass filters are connected in series to further amplify the DP signal. It is then mixed down to the intermediate frequency (IF) band using a local oscillator (LO) and mixer. After a final low-pass filter, the signal is digitized with an FPGA-based DAQ using a Xilinx RFSoC4x2, analogous to the system used in~\cite{BREAD:2023xhc}. The resulting IF spectrum spans from 0 to $\SI{2.4576}{\giga\hertz}$ with a resolution bandwidth of $\Delta\nu=\SI{9.375}{\kilo\hertz}$. The sensitive frequency range is set by the bandwidth of the narrowest band pass filter which results in an RF range of \SIrange{19.01}{20.30}{\giga\hertz}.

\begin{figure}
\centering
    \subfloat[\label{fig:bead_pull_long}]{%
        \centering%
        \includegraphics[width=\linewidth]{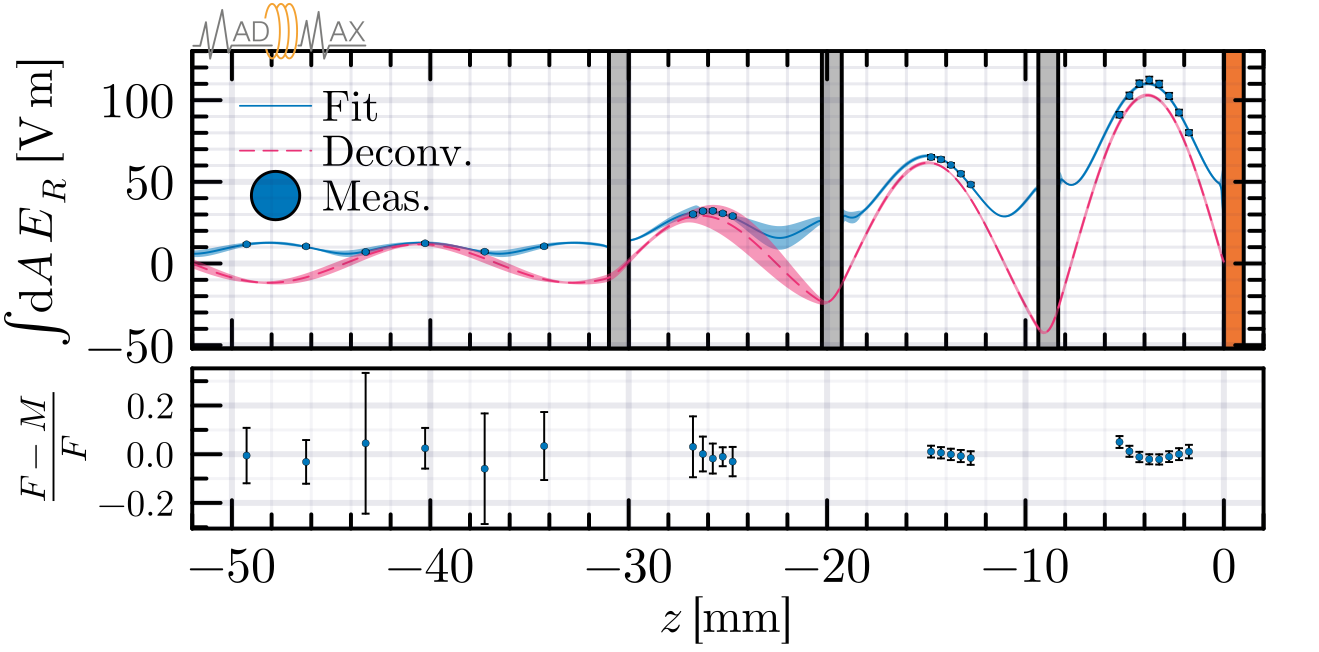}%
        }%
        
     \vspace{-5mm}  %
     \subfloat[\label{fig:bead_pull_boost}]{%
        \centering       %
        \includegraphics[width=\linewidth]{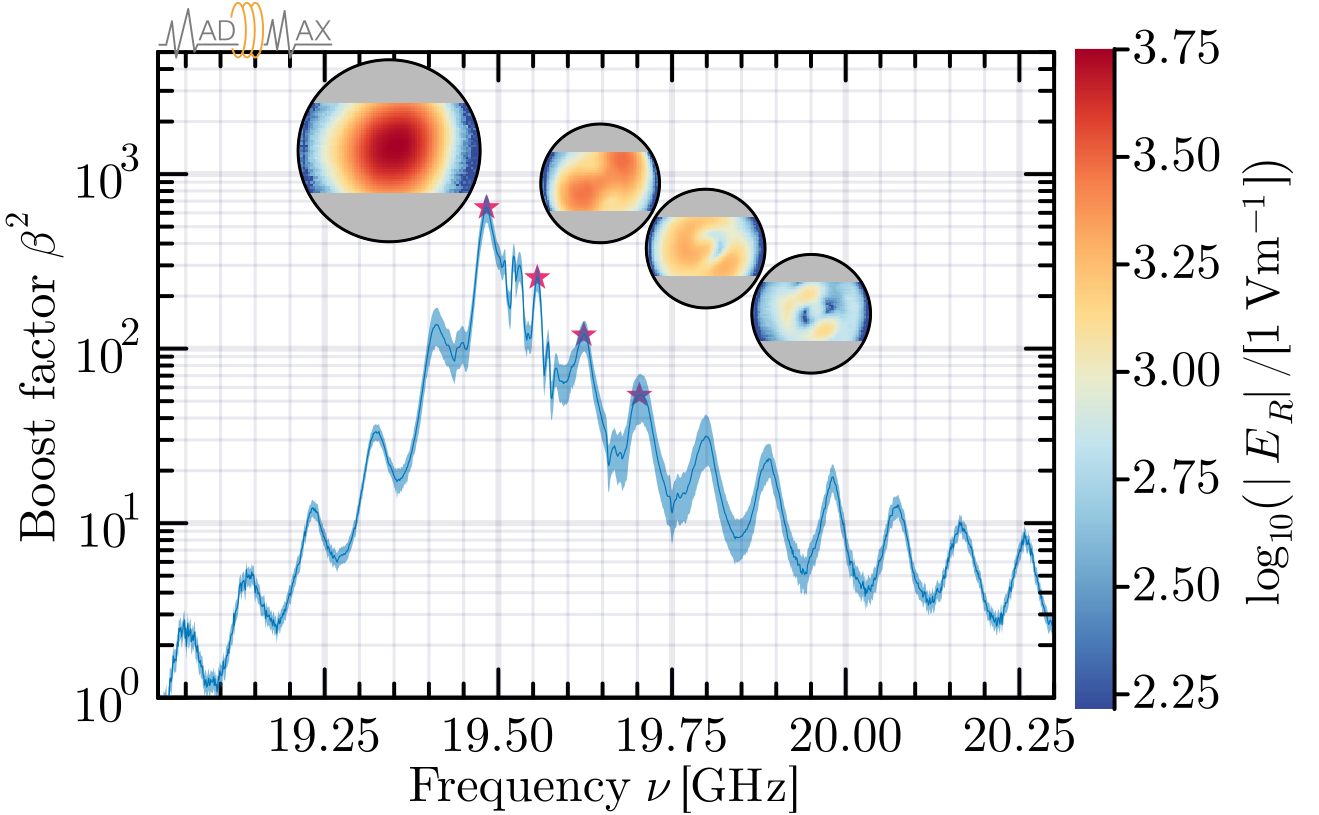}%
    }%
    \caption{Results of the boost factor determination using the bead-pull method. \textbf{(a)} Transverse integrated electric field along the optical axis inside the booster at \SI{19.48}{\giga\hertz}. Measurements (circles) fitted with a model (solid line) including the bead. The dashed line shows the field
    evaluated without the effect of the bead. The shaded band indicates model uncertainties from material and geometry parameters. The relative difference between fit (F) and measurement (M) is shown in the lower panel.
    \textbf{(b)} Boost factor as a function of frequency, including the corrections for the finite domain and receiver mismatch. The insets show the transverse electric field between the mirror and the first disk at the indicated frequencies (stars), where grey areas indicate regions unprobed due to mechanical constraints.}
    \label{fig:bead_pull}
\end{figure}

The expected DP signal power $P_{\rm{sig}}$ is proportional to the boost factor $\beta^2$. It can be determined directly from measurement using a recently developed method~\cite{Egge_2023} which relates the boost factor, primarily defined for the unknown DP-induced field, to the reflection-induced electric field ${E}_R$ that is excited by a Vector Network Analyzer (VNA). This field is measured using the non-resonant bead-pull method~\cite{steele1966} where small changes in the booster reflection coefficient $\Gamma$ are related to the electric field at the bead's position. The general procedure to determine $\beta^2$ from bead-pull measurements is described in detail in~\cite{Egge_2024,thesis_jacob_egge}. Expressed in terms of measurable quantities, the boost factor is
\begin{equation}
    \beta^2 = \frac{P_{\rm{sig}}}{P_0} = \frac{4\pi^2 \epsilon_0 \nu^2 F_{\rm{RC}}}{8 c P_{\rm{in}} A }\abs{\int \dd{V} {E}_R}^2,
    \label{Eq:boost_factor_reciprocity}
\end{equation}
where ${E}_R$ is excited with input power $P_{\rm{in}}$ at frequency $\nu$ and is spatially integrated over the conversion volume. $F_{\rm{RC}}$ is a dimensionless factor that accounts for an impedance mismatched receiver system. Bead-pull measurements yield ${E}_R$ relative to $P_{\rm{in}}$ such that its actual value is not required. We arbitrarily set $P_{\rm{in}}=\SI{1}{\watt}$ for convenience. Figure~\ref{fig:bead_pull} shows the longitudinal (a) and transverse field distribution (b) inside the booster obtained from bead-pull measurements as well as the derived boost factor as a function of frequency. 

The measured field is first integrated in the transverse direction. The absolute value of the integrated electric field is then fitted separately for each frequency by a model that takes the finite size of the bead into account, shown in Fig.~\ref{fig:bead_pull_long}. The shaded band indicates the model uncertainty from geometry and material parameters of both booster and bead. The fit allows us to obtain the deconvoluted field, i.e., the field without the response of the bead, to interpolate between measurements, and to recover the phase. The remaining integration in the longitudinal direction is then performed with the deconvoluted field, yielding $\beta^2$ via Eq.~\eqref{Eq:boost_factor_reciprocity}, shown in Fig.~\ref{fig:bead_pull_boost}. On the main resonance at $\SI{19.48}{\giga\hertz}$, the transverse field is the fundamental Gaussian mode with waist radius $w_0=\SI{91(6)}{\milli\meter}$ that is expected from the optical system and which has a good overlap with the uniform $\vb*{E}_{\chi}$. A mix of higher-order transverse modes can also resonate inside the booster as well as between the antenna and booster, causing the additional smaller peaks, first studied in simulation in~\cite{MADMAX_3D_2021}. 

The finite domain of the bead-pull measurements does not cover the full transverse extent of ${E}_R$ due to mechanical constraints from the booster leaving some of the top and bottom fringe areas of the disks unprobed (indicated by the grey areas in Fig.~\ref{fig:bead_pull_boost}). We account for these areas in the integral in Eq.~\eqref{Eq:boost_factor_reciprocity} by extrapolating the electric field. This increases the naive estimate of $\beta^2$ by about 70\%. The finite domain correction has been checked against independent measurements with full coverage of the field~\cite{thesis_jacob_egge} and contributes an additional 10\% to the systematic uncertainty of $\beta^2$. 

\begin{figure}
    \centering
    \includegraphics[width=\linewidth]{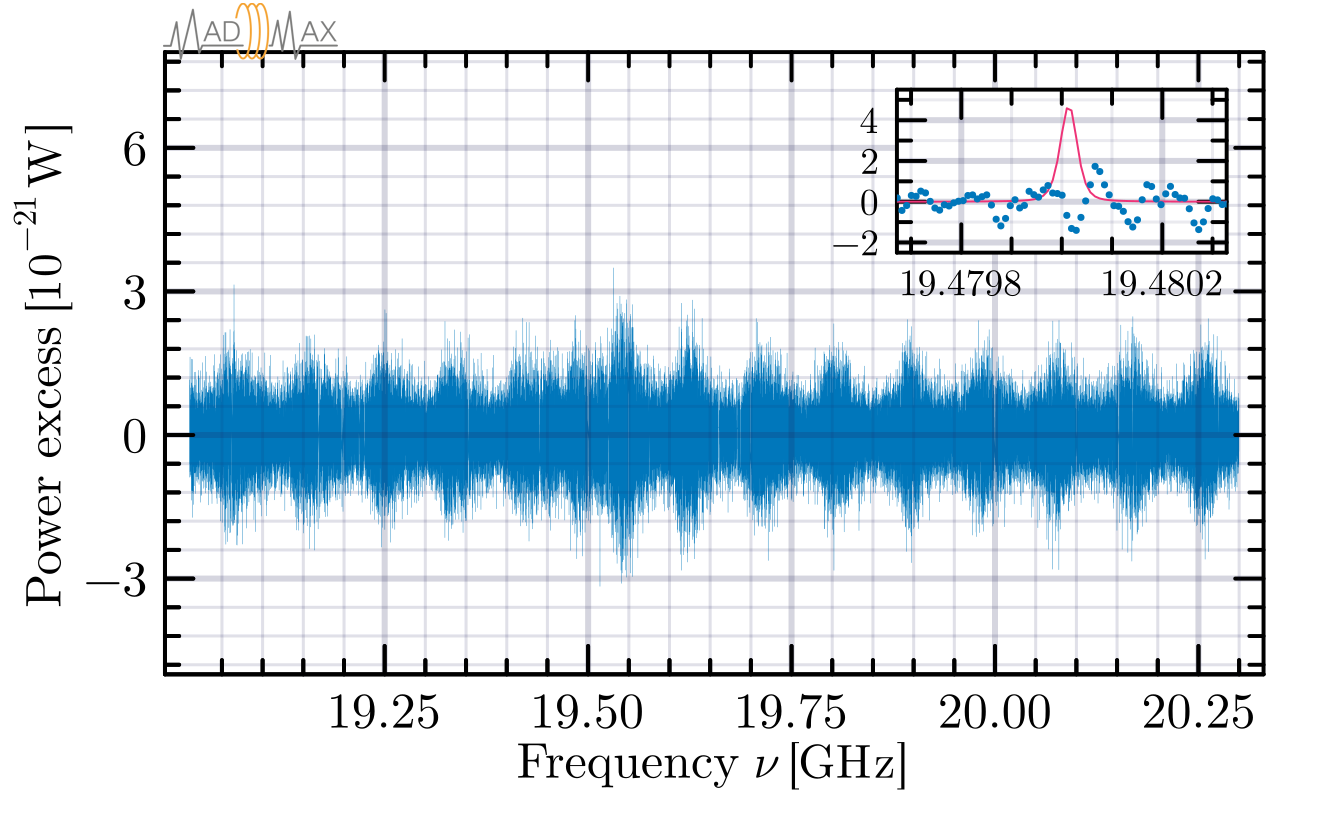}
    \caption{Observed cross-correlated power excess as a function of frequency. The inset shows a zoomed-in view around $\SI{19.48}{\giga\hertz}$ where the maximum boost factor occurs. A hypothetical DP signal with $\chi=2\times 10^{-13}$ (magenta solid line) is superimposed on observations (blue circles).}
    \label{fig:data}
\end{figure}

The receiver chain has an impedance mismatch to the antenna, resulting in an input reflection coefficient of $\abs{\Gamma_{\rm{RC}}} \sim 0.25$. Since part of the expected DP signal would be resonating between receiver chain and booster, the boost factor is further changed by the factor $F_{\rm{RC}}=\frac{1-\abs{\Gamma_{\rm{RC}}}^2}{\abs{1-\Gamma_{\rm{RC}} \Gamma}^2}$.
This resonance also affects the system noise temperature $T_{\rm{sys}}$ which is determined by a Y-factor calibration of the receiver chain~\cite{pozar2011microwave}, establishing its gain and equivalent noise temperature $T_e$. The overall system noise temperature appears as a standing-wave pattern with a peak-to-peak system temperature of $\SIrange{120}{332}{\kelvin}$ over the measurement span of $\SI{1.2}{\GHz}$, which is dominated by the receiver chain noise of $T_e = \SI{119 +- 28}{\kelvin}$ interfering with itself. This is compatible with the expected noise of the first stage amplifier of around $\SI{120}{\kelvin}$~\cite{lnf:lnc15_29b}. Only at the resonance frequency of the booster does its thermal radiation due to physical temperature significantly contribute to the system noise temperature by around 40\%. The modulation of $T_{\rm{sys}}$ is visible in the received power excess in Fig.~\ref{fig:data} where a higher $T_{\rm{sys}}$ leads to larger fluctuations around the baseline. 

To properly quantify $F_{\rm{RC}}$, the difference in electrical length between the measurements of $\Gamma$ and $\Gamma_{\rm{RC}}$ needs to be known. It is extracted from fitting a 1D model to the system noise temperature, taking $\Gamma_{\rm{RC}}$ and $\Gamma$ as input and simulating the emitted LNA noise as per~\cite{twoportnoise}. The resulting value of $\SI{16.016(2)}{\mm}$ matches the length of the adapter used to connect the receiver chain to the antenna. Correcting for the mismatch increases the systematic uncertainty of $\beta^2$ by less than 1\%. Depending on frequency, $F_{\rm{RC}}$ ranges between 0.6 and 1.6, and can consequently increase or decrease the boost factor. By tuning the distance between the antenna and booster, it is assured that on resonance frequency the boost factor is increased by $F_{\rm RC}\approx 1.3$ to a maximum value of $\max \beta^2=\num{640(110)}$. The boost factor shown in Fig.~\ref{fig:bead_pull_boost} includes both corrections from the finite domain of bead-pull measurements and receiver chain mismatch. The shaded band indicates the one standard deviation uncertainty which ranges from 10 to 36\%, depending on frequency.

{\bf\textit{Data Taking Run.}}---The data-taking run lasted 16.5 days from 2023-12-22 to 2024-01-08. In this period, ${N_{\rm{tot}}=\num{9.484e9}}$ spectra were measured corresponding to an effective data-taking time of $\Delta t = \SI{11.7}{\day}$. The spectra were averaged in batches of $N_{\rm{av}}=\num{4.00e6}$ and saved every 10 minutes resulting in 2371 files. The setup was located in a shielded laboratory (SHELL) at DESY/University of Hamburg with coordinates: 53.58N/9.89E. The antenna is sensitive only to the zenith-pointing polarization of DPDM. The optical axis of the booster has an azimuth angle of 65° relative to north. 

The data-taking routine employed the same LO hopping scheme as described in~\cite{BREAD:2023xhc} to smear and suppress radio frequency interference (RFI) in the IF band. Approximately once per second (every $10^4$ spectra), the LO frequency is randomly shifted within a window of $\sim\SI{9}{\mega\hertz}$. The collected power spectra are then realigned in the RF band before averaging. This results in a smearing of any RFI in the IF band, effectively suppressing it without affecting signals in the RF band. Additionally, roughly 50 bins that contain strong RFI signals in the IF band are digitally masked. The frequency hopping also causes these masks to spread in the RF band, rendering their impact on sensitivity negligible. 

Communication delays between DAQ and LO are the dominant contribution to the dead time of $\sim$ 30\% and could be improved in the future. In the RF band, RFI signals are further suppressed by $\sim\SI{30}{\deci\bel}$ using an additional Faraday cage and RF absorbers. The response of these signals to additional shielding excludes a DP origin. They are removed from the analysis and are listed in the supplementary materials. A plausible source is the surface-monitoring radar of the nearby Hamburg airport.

\begin{figure*}
    \centering
    \includegraphics[width=\textwidth]{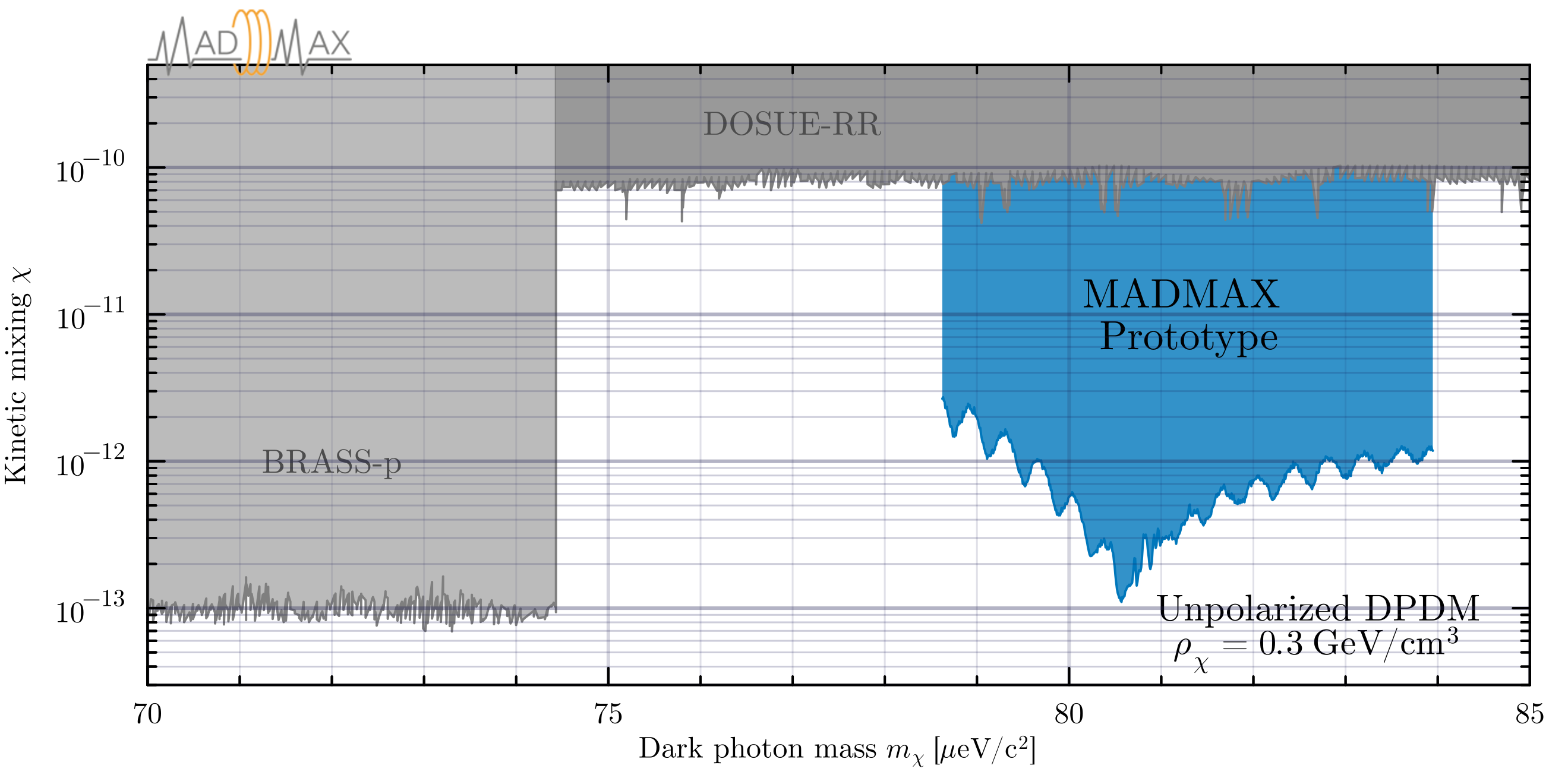}
    \caption{95\% CL upper limit on the dark photon kinetic mixing angle $\chi$ obtained with the \MADMAX{} prototype as compared to the dish antenna experiments DOSUE-RR~\cite{DOSUE-RR:2022ise} and BRASS-p~\cite{brass2023}, rescaled to a common value of the assumed local dark photon dark matter density of $\rho_{\chi}=\SI{0.3}{\giga\eV\per\cm^3}$. Unpolarized dark photons are assumed.}
    \label{fig:limit}
\end{figure*}

{\bf\textit{Analysis.}}---The analysis procedure and nomenclature closely follows HAYSTAC~\cite{haystac2017}. All saved spectra $P_{i}(\nu)$ are each filtered with a 6th-order Savitzky-Golay (SG) filter with a window length of $\SI{1.04}{\mega\hertz}\gg\Delta\nu_{\chi}$ that would leave a potential DP signal intact. From this we obtain baselines $P_{{\rm bl},i}(\nu)$ and processed spectra $p_{{\rm proc},i}(\nu) = (P_{i}/P_{{\rm bl},i} - 1)$. For perfect baseline subtraction, the standard deviation $\sigma_i$ of each processed spectrum, excluding RFI bins, is expected to be $1/\sqrt{N_{\rm{av}}}$. We observe $\sigma_i$ to be between 97.9\% and 99.2\% of this value, indicating the baseline subtraction has a minor impact on our sensitivity. The SG filter reduces the SNR of a potential DP signal by $\eta_{\rm SG}=0.91$ which is obtained by running the analysis on simulated data with injected synthetic DP signals. During the run, the baseline drifted by no more than 5\% in amplitude which can be attributed to gain variation caused by room temperature changes of the same order. In frequency, spectral features of $P_{{\rm bl},i}$ shifted by less than $\SI{1}{\mega\hertz}$. Assuming the same frequency stability for $\beta^2$, which was determined after the run, translates to a relative systematic uncertainty of $\sim$ 4\% in amplitude. 

The processed spectra are combined using weights that take the expected ${\rm{SNR}}$ of a hypothetical single-bin DP signal into account for every bin of each spectrum~\cite{haystac2017}. The resulting combined spectrum is further cross-correlated with the expected DP line shape~\cite{DIEHL2024169259,ohare2017_boost_max_boltz} to account for the fact that the expected DP signal would stretch over $\sim$ 5~bins. We use $\sigma_v=\SI{154.1}{\kilo\metre\per\second}$~\cite{Bovy_2012_sigma_v} for the velocity dispersion of the local DM halo and $v_{\rm{lab}}=\SI{242.1}{\kilo\metre\per\second}$~\cite{malhan2020_v_lab} for the velocity of the laboratory with respect to the local standard of rest. The unknown relative alignment between the frequency bins and DP line shape leads to variation in the discretization of the line shape, affecting the cross-correlation, and is treated as a systematic uncertainty. The observed cross-correlated power excess $P_{\rm{exc}}$ is shown in Fig.~\ref{fig:data} revealing a sensitivity to DP signals of $\sim\SI{2e-21}{\watt}$. A hypothetical DP signal with $\chi=2\times 10^{-13}$ is shown in the inset along with a zoomed-in view of the observations around \SI{19.48}{\giga\hertz} where the maximum boost factor occurs. 

\begin{table}
\caption{\label{tab:systematics} Summary of systematic uncertainties. Min-to-max is shown for frequency-dependent quantities.}
\vspace{0.2cm}
\begin{tabular}{@{}lr@{}}
\toprule
Effect                                  & Uncertainty on $\chi$ \\ \midrule
Boost factor determination & \\
~~~Bead-pull measurements & 2 to 17\%\\
~~~Bead pull finite domain correction & 5\%\\
~~~Receiver chain impedance mismatch & $<$1\%\\
Subtotal & 5 to 18\% \\
\midrule
Y-factor calibration & 4\%  \\ 
Power stability & 3\%\\
Frequency stability & 2\%\\
Line shape discretization & 4\%\\ 
\midrule
Total & 9 to 19\%\\
\bottomrule
\end{tabular}
\end{table}

The largest excess has a local significance of $4.3 \sigma$. This is within the expectation of observing thermal noise only as the probability of observing a local excess at least this large is $p \approx 0.21$ for the full dataset. From the non-observation of any DP signal, we derive a 95\% confidence level (CL) upper limit on $P_{\rm{exc}} \rightarrow P_{\rm{exc}}^{95}$ for groups of 210 neighboring bins using the approach described in~\cite{ADMX:2020hay}. The limit on the kinetic mixing angle $\chi$ is obtained by setting $P_{\rm{exc}}^{95} = \eta_{\rm SG} \beta^2 P_0$. Systematic uncertainties, summarized in Table~\ref{tab:systematics}, are assumed to be independent and propagated to $\chi$. The most conservative value of $\chi$ within this uncertainty is adopted for the final limit.

{\bf\textit{Conclusion.}}---The 95\% CL limit on the dark photon kinetic mixing angle $\chi$ is shown in Fig.~\ref{fig:limit}. Unpolarized dark photon dark matter ($\rho_{\chi}=\SI{0.3}{\giga\eV\per\cm^3}$) with $\chi> \num{2.7e-12}$ can be excluded for masses between $\SIrange{78.62}{83.95}{\micro\eVperc^2}$. Using only three disks, we have improved existing limits by up to almost three orders of magnitude with a peak sensitivity of $\chi = \num{1.1e-13}$. This demonstrates, for the first time, the feasibility of the \MADMAX{} dielectric haloscope concept and show-cases its capability of providing a large conversion volume at high frequencies while retaining resonant enhancement. Further improvements are planned: scaling the conversion volume by adding more dielectric disks, reducing system noise temperature using a cryostat, and implementing a tunable booster via motorized disk control~\cite{MADMAX_piezo_2023,madmax_p200}. This will vastly improve both the mass range of searches for dark photons as well as increase the sensitivity to their coupling to photons.

\begin{acknowledgments}
{\bf\textit{Acknowledgments.}}
We thank Olivier Rossel for providing the drawing of the setup and artistic support. A big thanks to the BRASS~\cite{brass2023} group and in particular Le Hoang Nguyen who provided us with vital equipment. We thank Xiaoyue Li for her contributions to understanding dielectric haloscopes and helpful discussions. This work is supported by the Deutsche Forschungsgemeinschaft (DFG, German Research Foundation) under Germany’s Excellence Strategy, EXC 2121, Quantum Universe (390833306). We acknowledge the support of the \MADMAX{} project by the Max Planck Society. S.K., G.C., M.F., L.S. are supported by Fermi Research Alliance, LLC under Contract No. DE-AC02-07CH11359 with the U.S. Department of Energy, Office of Science, Office of High Energy Physics. The work of J.R. is supported by Grants PGC2022-126078NB-C21 funded by MCIN/AEI/ 10.13039/501100011033 and “ERDF A way of making Europe” and Grant DGA-FSE grant 2020-E21-17R Aragon Government and the European Union - NextGenerationEU Recovery and Resilience Program on ‘Astrof\'isica y F\'isica de Altas Energ\'ias’ CEFCA-CAPA-ITAINNOVA. G. H. would like to thank the University of Chicago Joint Task Force Initiative for its generous support for this research.
\end{acknowledgments}

\bibliographystyle{JHEP}
\bibliography{references}

\onecolumngrid
\appendix

\clearpage

\setcounter{equation}{0}
\setcounter{figure}{0}
\setcounter{table}{0}
\setcounter{page}{1}
\makeatletter

\begin{center}
\textbf{\large Supplemental Material for the Letter\\[0.5ex]
{\em First search for dark photon dark matter with a MADMAX prototype}}
\end{center}

This Supplemental Material includes additional figures of relevant quantities not shown in the main text as well as a list of frequencies that are excluded from the final limit due to radio interference. Figures are adapted from~\cite{thesis_jacob_egge}.

\FloatBarrier

\begin{figure*}
    \centering
    \includegraphics[width=0.75\linewidth]{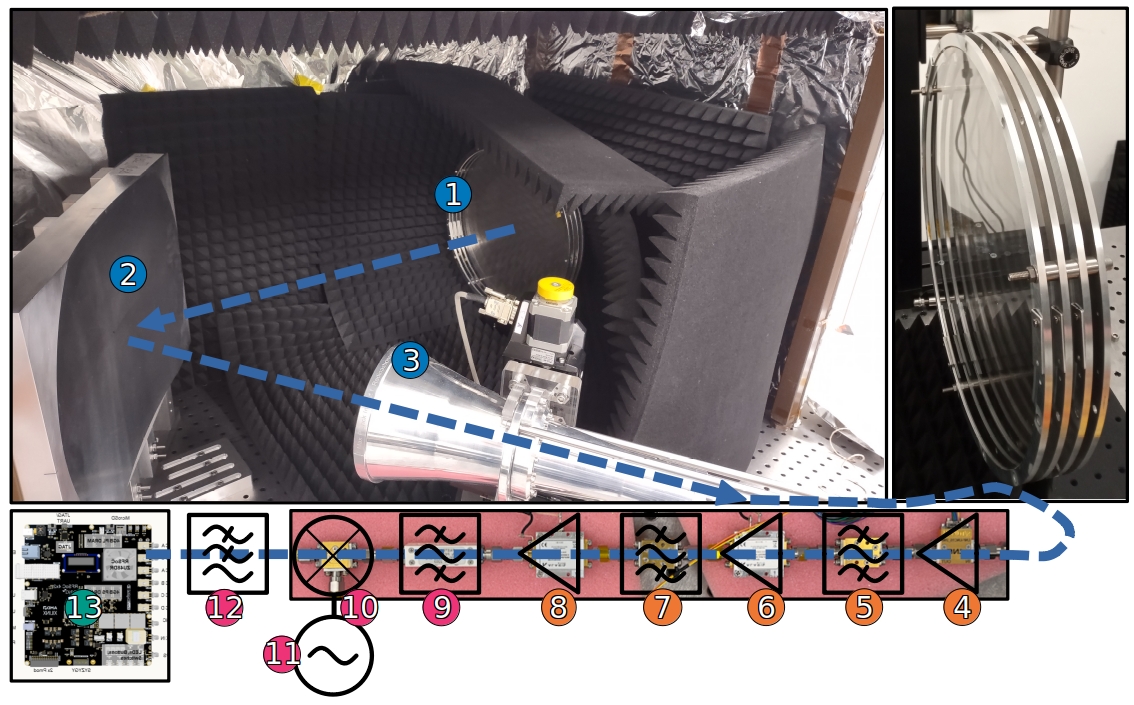}
    \caption{The MADMAX prototype setup in the shielded laboratory SHELL at DESY/University of Hamburg. The dashed line indicates the path of a putative dark photon signal. The individual components are: 1) Booster consisting of three dielectric disks and a metallic mirror with $\SI{30}{\centi\metre}$ diameter. 2) Focusing mirror. 3) Horn antenna. 4,6,8) Low noise amplifiers. 5,7,9) Bandpass filters. 10) Mixer. 11) Local oscillator. 12) Lowpass filter. 13) DAQ. A Faraday cage and RF absorbers provide additional shielding against external interference.}
    \label{fig:setup_pic}
\end{figure*}

\begin{figure*}
    \centering
    \includegraphics[width=0.5\linewidth]{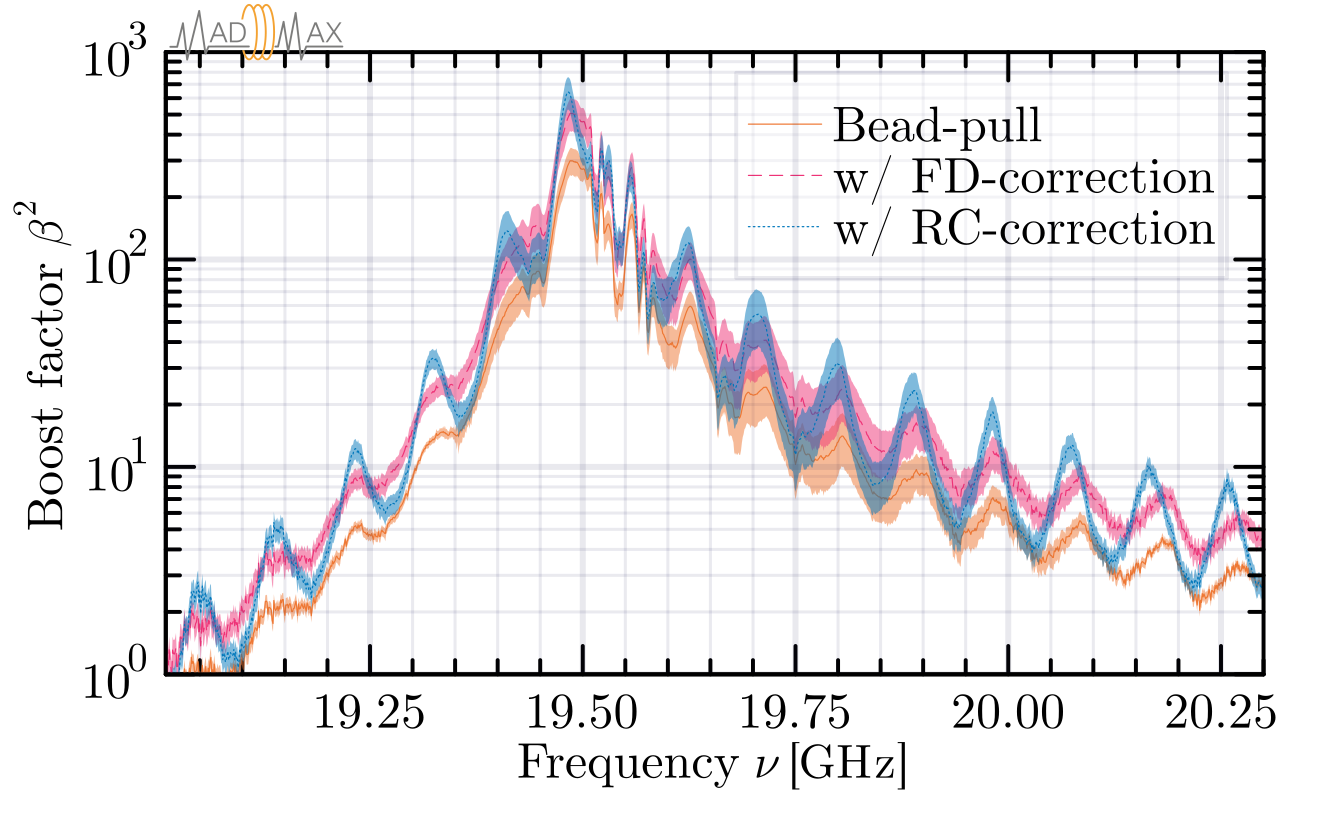}
    \caption{Boost factor with and without corrections as a function of frequency. The orange solid line shows the result as obtained from bead-pull measurements that neither cover the full extent of the booster nor include the receiver chain. Correcting for the finite domain (FD) of the bead-pull measurements increases the boost factor by $\sim 70\%$ (magenta dashed line). The receiver chain (RC) has a slight mismatch to the antenna and thus reflects part of the signal back into the booster. This modulates the boost factor by a factor $F_{\rm RC}$ ranging from 0.6 to 1.6. The blue dotted line shows the boost factor with both finite domain and receiver chain correction. The shaded band indicates the uncertainty coming from measurement, geometry, and material parameters.}
    \label{fig:bf_fd_rc_comp}
\end{figure*}

\begin{figure*}
    \centering
    \includegraphics[width=0.5\linewidth]{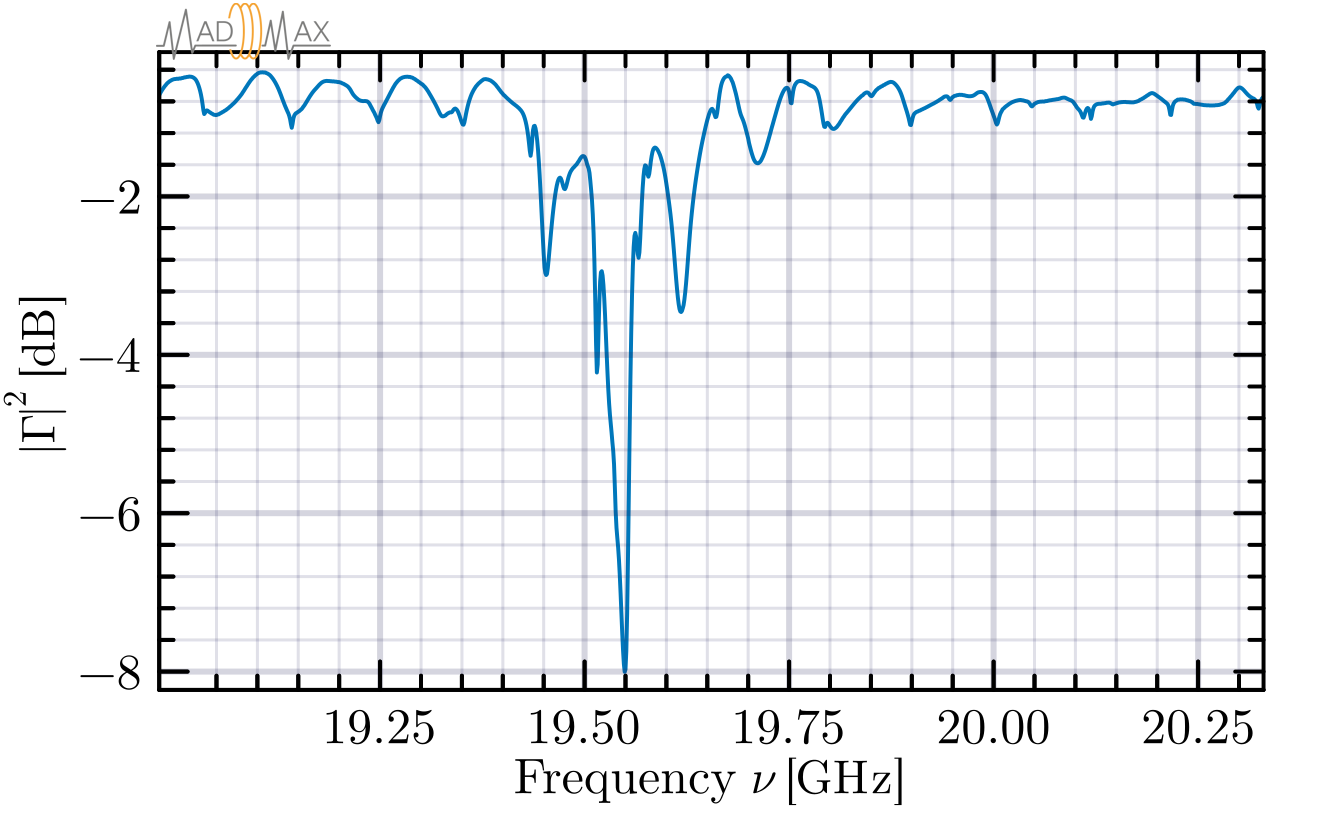}
    \caption{Booster reflection coefficient $\Gamma$ as a function of frequency. The booster resonance manifests itself as a broad dip around $\SI{19.5}{\giga\hertz}$ that is superimposed by a standing wave pattern due to reflections between the antenna and booster. The sharp dip at $\SI{19.55}{\giga\hertz}$ is mainly caused by the latter.}
    \label{fig:Gamma}
\end{figure*}

\begin{figure*}
    \centering
    \includegraphics[width=0.5\linewidth]{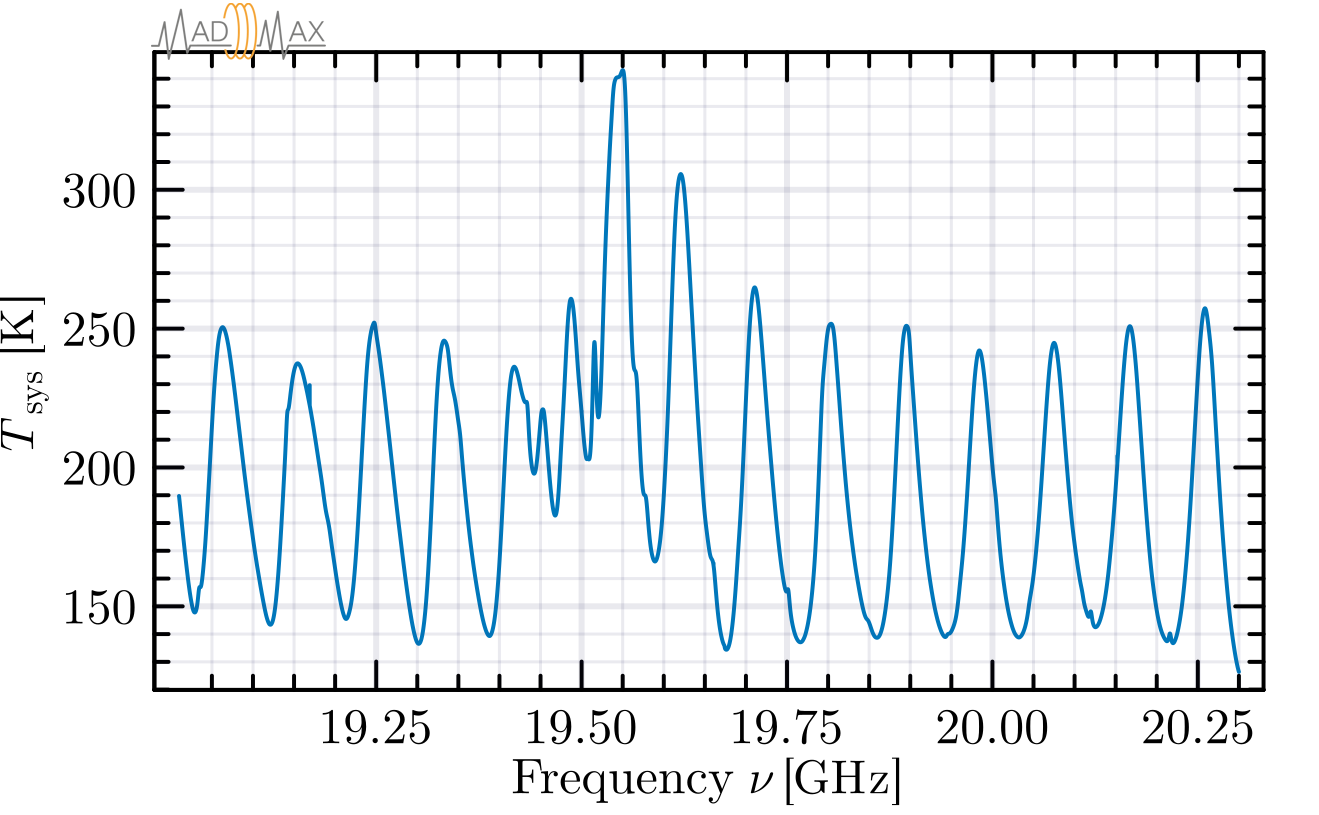}
    \caption{System temperature $T_{\rm sys}$ of the booster and receiver system as a function of frequency. The standing wave pattern is caused by the added noise of the first stage amplifier that is reflected by the booster and interfering with itself. The booster resonance around $\SI{19.48}{\giga\hertz}$ increases the optical path length between the first stage amplifier and booster, squeezing the standing wave pattern in frequency. Thermal radiation from the booster only becomes relevant where its reflection coefficient $\Gamma$ is small and causes the peak in $T_{\rm sys}$ at $\SI{19.55}{\giga\hertz}$.  }
    \label{fig:T_sys}
\end{figure*}

\begin{figure*}
    \centering
    \includegraphics[width=0.5\linewidth]{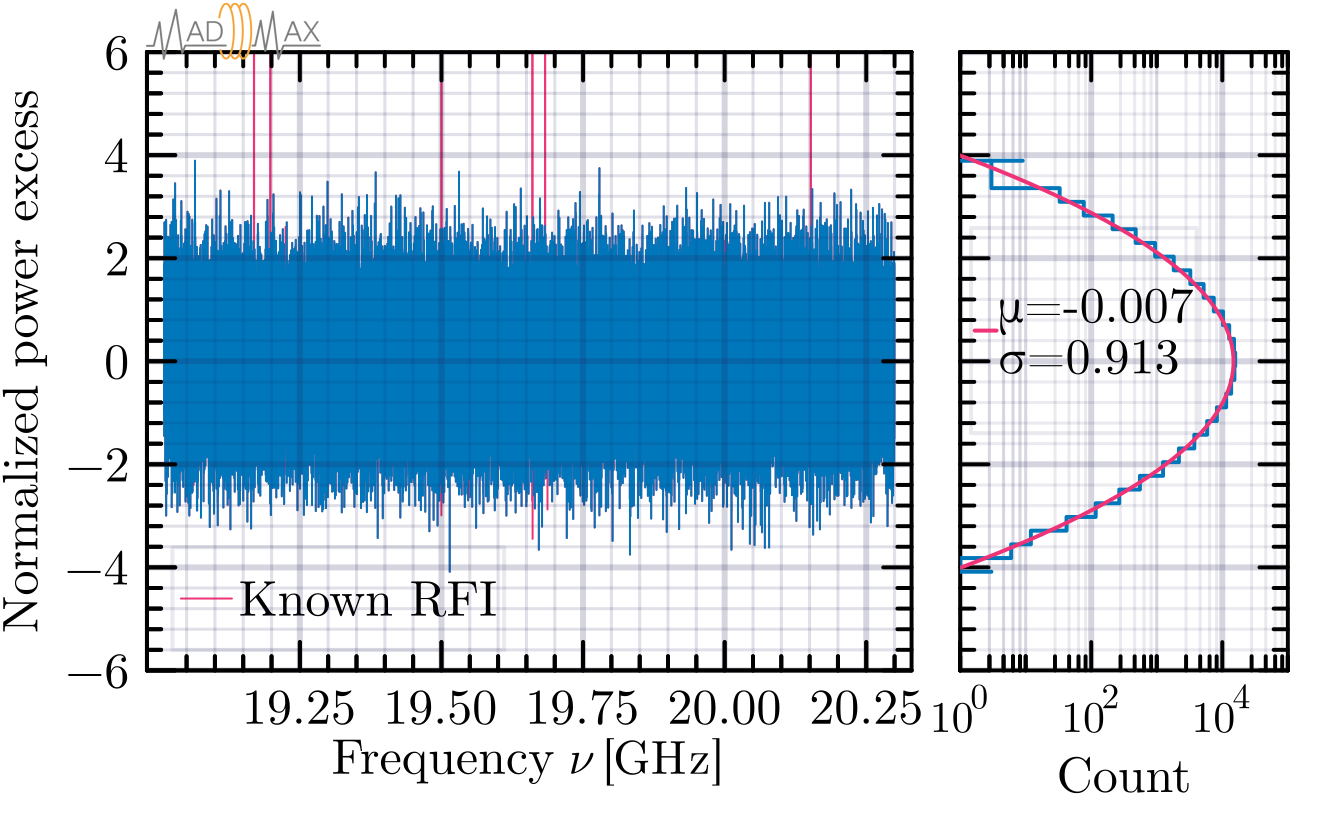}
    \caption{Normalized cross-correlated power excess, also known as a grand spectrum as a function of frequency. Known RFI peaks are shown in magenta and are excluded from the analysis. The remaining data is Gaussian distributed, as shown in the right projection. The standard deviation below unity is due to the correlation between Savitzky-Golay filter and the DP line shape. Monte Carlo simulations of thermal noise yield an expected value of $\sigma=0.906$, close to the observed value of $0.913$.}
    \label{fig:grand_spectrum}
\end{figure*}

\begin{table*}[]
    \caption{List of RFI peaks observed in an unshielded rejection spectrum with a local significance $>7\sigma$. All RFI signals could be attenuated by $\sim \SI{30}{\decibel}$ using additional shielding to the point where only 7 peaks show up in the shielded analysis run above the noise floor of around $-182$ dBm. Nonetheless, all listed frequencies are excluded from the analysis and thus from the final limit.}
    \vskip 12pt
    \centering
    \begin{tabular}{l l c c l l}
         Peak & Frequency [GHz] & \multicolumn{2}{l|}{Excess power [dBm]} & \multicolumn{2}{l|}{Local significance}\\
         \hline
  &                   & Unshielded& Shielded & Unshielded& Shielded \\ 
\hline
1  & 19.06198-19.06261 & $-166$ & $-181$ & 14 & 3\\
2  & 19.16860-19.16956 & $-119$ & $-150$ & 627838 & 3181\\
3  & 19.19670-19.19733 & $-159$ & $-180$ & 74 & 5\\
4  & 19.19755-19.19820 & $-153$ & $-178$ & 288 & 8\\
5  & 19.19948-19.20011 & $-165$ & $-182$ & 17 & 3\\
6  & 19.21823-19.21886 & $-164$ & $-183$ & 32 & 3\\
7  & 19.22462-19.22524 & $-170$ & $-183$ & 8 & 3\\
8  & 19.35308-19.35371 & $-167$ & $-181$ & 13 & 2\\
9  & 19.43971-19.44033 & $-166$ & $-181$ & 14 & 3\\
10 & 19.49948-19.50038 & $-132$ & $-160$ & 32118 & 324\\
11 & 19.52462-19.52524 & $-163$ & $-181$ & 36 & 2\\
12 & 19.54962-19.55025 & $-167$ & $-179$ & 9 & 2\\
13 & 19.57461-19.57525 & $-169$ & $-182$ & 10 & 2\\
14 & 19.66013-19.66106 & $-127$ & $-160$ & 175881 & 499\\
15 & 19.68271-19.68334 & $-170$ & $-178$ & 9 & 9\\
16 & 19.68698-19.68761 & $-170$ & $-182$ & 9 & 2\\
17 & 19.69948-19.70012 & $-161$ & $-182$ & 64 & 2\\
18 & 19.99948-20.00034 & $-161$ & $-181$ & 46 & 3\\
19 & 20.09948-20.10011 & $-168$ & $-182$ & 8 & 3\\
20 & 20.15169-20.15254 & $-135$ & $-155$ & 28471 & 1272\\
21 & 20.16872-20.16935 & $-170$ & $-180$ & 7 & 3\\
22 & 20.27468-20.27531 & $-163$ & $-181$ & 29 & 3\\

    \end{tabular}
    \label{tab:rfi_bins}
\end{table*}

\end{document}